\documentclass[aps,twocolumn]{revtex4}
\makeatletter

\newcommand{\Rmnum}[1]{\expandafter\@slowromancap\romannumeral #1@}
\makeatother
\usepackage{graphicx}
\usepackage{dcolumn}
\usepackage{bm}
\usepackage{CJK}
\usepackage{color}
\usepackage{amsfonts}
\usepackage{psfrag}
\usepackage{wrapfig}
\usepackage{subfigure}
\usepackage{makeidx}
\usepackage{multirow}
\usepackage{epsf}
\usepackage[colorlinks,linkcolor=red,anchorcolor=red,citecolor=blue,urlcolor=blue]{hyperref}
\usepackage{amsmath}
\usepackage{cases}
\usepackage{bm}

\begin{document}
\begin{CJK}{GBK}{song}
\title{Chess-Board-Like Spatio-Temporal Interference Patterns and Their Excitation}
\author{Chong Liu$^{1,2,3}$}
\author{Zhan-Ying Yang$^{1,3}$}
\author{Wen-Li Yang$^{1,3,4}$}
\author{Nail Akhmediev$^{2}$}
\address{$^1$School of Physics, Northwest University, Xi'an 710069, China}
\address{$^{2}$Optical Sciences Group, Research School of Physics and Engineering, The Australian National University, Canberra, ACT 2600, Australia}
\address{$^3$Shaanxi Key Laboratory for Theoretical Physics Frontiers, Xi'an 710069, China}
\address{$^4$Institute of Modern Physics, Northwest University, Xi'an 710069, China}

\begin{abstract}
We discover new type of interference patterns generated in the focusing nonlinear Schr\"odinger equation (NLSE) with localised periodic initial conditions. At special conditions, found in the present work, these patterns exhibit novel chess-board-like spatio-temporal structures which can be observed as the outcome of collision of two breathers.
The infinitely extended chess-board-like patterns correspond to the continuous spectrum bands of the NLSE theory.
More complicated patterns can be observed when the initial condition contains several localised periodic swells.
These patterns can be observed  in a variety of physical situations ranging from optics and hydrodynamics to Bose-Einstein condensates and plasma.
\end{abstract}


\maketitle

The concept of solitons is well known and it was proven to be useful in studies of tsunami waves, pulses in optical fibres and matter waves in Bose-Einstein condensate.
Large part of theoretical studies on solitons is based on the Korteweg-de Vries and the nonlinear Schr\"odinger equations (NLSE).
The latter describes multiplicity of phenomena in various branches of physics and can be considered as an archetypical model of nonlinear wave propagation.  In addition to solitons, the NLSE has also a family of breather solutions.
In contrast to ordinary solitons, breather solutions are not localised. They are located on top of a plane wave.
Breathers have been studied in hydromechanics \cite{W1a,W1b}, nonlinear optics \cite{O1a,O1b}, optomechanics \cite{OM1,OM2}, plasma \cite{plasma}, metamaterials \cite{MM}, and Bose-Einstein condensates \cite{BEC}. They play an important role in describing variety of physical phenomena such as modulation instability (MI) \cite{Akh88,Wabnitz,Erkintalo,SR1}, supercontinuum generation \cite{SCG}, turbulence \cite{Agafon,Suret,Suret1,IT3a,IT3b}, Fermi-Pasta-Ulam recurrence \cite{FPU0,FPU1,FPU2,FPU3,FPU4}, Talbot effect \cite{NT}, and rogue wave events \cite{RW1}. Recently, significant progress has been made on the experimental observation of fundamental breathers \cite{FB1,FB2,FB20,FB3a,FB3b} and their nonlinear superpositions \cite{HB1,HB2,HB3}, both in nonlinear optics and hydrodynamics.  Just as solitons, breathers may collide with each other creating complex interference patterns. They reveal high amplitude peaks when the breathers are synchronised \cite{RW1,BI,BI1}, similar to the case of the soliton synchronisation \cite{SI1a,SI1b}.

One way of creating breathers in nonlinear physics is through MI.
The latter is the exponential growth of periodic perturbations of a plane wave in unstable media \cite{PP}.
It is known as Benjamin-Fair instability \cite{BF} in the case of water waves or Bespalov-Talanov instability \cite{Bespalov} in the case of optical beams. The NLSE has an exact solution that describes both the process of the initial growth and the following full scale recurrent evolution back to the plane wave. These solutions have been first given in \cite{B01,AB1} and presently known as `Akhmediev breathers' (AB). A single breather produces localised periodic fringe pattern.
Fringes are also produced when two solitons collide (soliton interference pattern) \cite{SI-1,SI-2,SI-3,SI-4}. On the contrary, collision of two breathers can produce new spatio-temporal patterns which are doubly periodic. This is an important difference between solitons and breathers that deserves special attention.

Before entering the details of this phenomenon let us recall that MI can be produced using localised periodic perturbations. Analytic solutions for this phenomenon are known as super-regular breathers \cite{SR1,HB3,CL}. The latter is the result of collision of two ABs propagating at small angles of opposite sign to the direction of evolution. It has been found \cite{SR1} that modulation in this case also grows producing periodic fringes just like in the case of ordinary MI. Modulation remains localised although the size of localisation increases in evolution. However, this happens only if the period of modulation remains within the instability interval. If the period is chosen outside of this interval, the evolution pattern changes drastically. In the present work, we found that the pattern transforms itself into a chess-board-like structure. Moreover, expanding this pattern leads to a continuous spectrum of eigenvalues involved in the NLSE theory. The latter is a nontrivial result that may significantly influence our vision of the role of breathers in nonlinear wave evolution.

We start with the focusing dimensionless NLSE \cite{Book97}:
\begin{eqnarray}
i\frac{\partial\psi}{\partial z}+\frac{1}{2} \frac{\partial^2\psi}{\partial t^2}+|\psi|^2\psi=0,\label{eq1}
\end{eqnarray}
where $\psi(z,t)$ is the wave envelope, $z$ is the propagation variable, and $t$ is the transverse variable.
This equation governs the nonlinear wave evolution in various media.
In particular, it describes gravity waves in deep-water conditions \cite{W1a} and
light waves in optical fibers \cite{O1a}. In each case, the function $\psi(z,t)$ describes the envelope of
the modulated waves and its absolute value carries information about either wave elevation above the average water surface or
the intensity of optical waves.

Let us consider the simplest case of symmetric collision of two breathers with the same amplitude and opposite group and phase velocities.
This exact solution can be constructed using Darboux transformation  with a pair of complex eigenvalues $\lambda_j$ with equal real parts and equal imaginary parts of opposite sign \cite{B01}.
The solution can be generalised for arbitrary number of breathers \cite{Akh88}.
The eigenvalues can be parameterised using the Joukowsky (Zhukovsky) transform \cite{SR1}. Namely, $\lambda_j=i(\xi+1/\xi)/2$, with $\xi=re^{\pm i\alpha}$, i.e., $\lambda_j=(i\nu\mp\mu)/2$, with $\nu=\epsilon_+\cos \alpha$, $\mu=\epsilon_-\sin \alpha$, $\epsilon_\pm=r\pm1/r$. Here $r~(\geq1)$ and $\alpha$ are the radius and the angle on the plane of polar coordinates that determine the value of $\lambda_j$. This transformation provides
certain convenience in presentation of the condition of chess-board-like interference patterns.

The two-breather solution can be written in the form:
\begin{equation}
\psi(z,t)=\left[1-\frac{G(z,t)+iH(z,t)}{D(z,t)}\right]\exp(iz),\label{eqb}
\end{equation}
where $G$, $H$, and $D$ are real functions of two variables
\begin{eqnarray}
G&=&\nu \mu \big\{4\nu\mu \left[ \cosh (\bm{\kappa}_1-\bm{\kappa}_2)+\cos (\bm{\phi}_1+\bm{\phi}_2) \right]\nonumber\\
&&+\nu A_2 \Delta_1-\mu A_1\Delta_2 \big\},\\
H&=&\nu \mu\bigg[2 \left(r^2-\frac{1}{r^2} \right)\cos2\alpha\sinh(\bm{\kappa}_1-\bm{\kappa}_2)+\gamma B_1 \Xi_1\nonumber\\
&&-2 \left(r^2+\frac{1}{r^2} \right)\sin2\alpha\sin(\bm{\phi}_1+\bm{\phi}_2)+\delta B_1\Xi_2 \bigg],\\
D&=&-\frac{1}{2} \gamma^2B_1\cos (\bm{\phi}_1-\bm{\phi}_2)-2 \sin ^22\alpha\cos (\bm{\phi}_1+\bm{\phi}_2)\nonumber\\
&&-\frac{1}{2} \nu^2A_2\sinh \bm{\kappa}_1\sinh \bm{\kappa}_2-2\nu\mu \left(\mu\Delta_2+\nu\Delta_1 \right)\nonumber\\
&&+ \left( \frac{1}{2} \nu^2B_2+\epsilon_+^2 \mu^2 \right)\cosh\bm{\kappa}_1\cosh\bm{\kappa}_2,
\end{eqnarray}
with
$\Delta_1=\sin \bm{\phi}_1 \sinh \bm{\kappa}_2-\sin \bm{\phi}_2 \sinh \bm{\kappa}_1,~
\Delta_2=\cos \bm{\phi}_1\cosh \bm{\kappa}_2+\cos \bm{\phi}_2 \cosh \bm{\kappa}_1,~
\Xi_1=\cos \bm{\phi}_1\sinh \bm{\kappa}_2-\cos \bm{\phi}_2 \sinh \bm{\kappa}_1,~
\Xi_2=\sin \bm{\phi}_1\cosh \bm{\kappa}_2+\sin \bm{\phi}_2\cosh \bm{\kappa}_1$, $\gamma=\epsilon_- \cos \alpha$, $\delta=\epsilon_+ \sin \alpha$,
$A_j=\epsilon_+^2+2 \cos2 \alpha\pm2$, $B_j=r^2+1/r^2\pm2\cos2\alpha$.
Here, $\bm{\kappa}_1=\gamma(t-V_{gr} z)$, $\bm{\kappa}_2=\gamma(t+V_{gr} z)$, $\bm{\phi}_1=\delta(t-V_{ph} z)+\theta_1$,
$\bm{\phi}_2=-\delta(t+V_{ph} z)+\theta_2$, $V_{gr}=-(\delta\nu/\gamma+\mu)/2$, $V_{ph}=(\gamma\nu/\delta-\mu)/2$.
Here, $V_{gr}$ and $V_{ph}$ are the group and phase velocities
while $\theta_j$ are arbitrary phases.

\begin{figure}[htb]
\centering
\includegraphics[height=60mm,width=84mm]{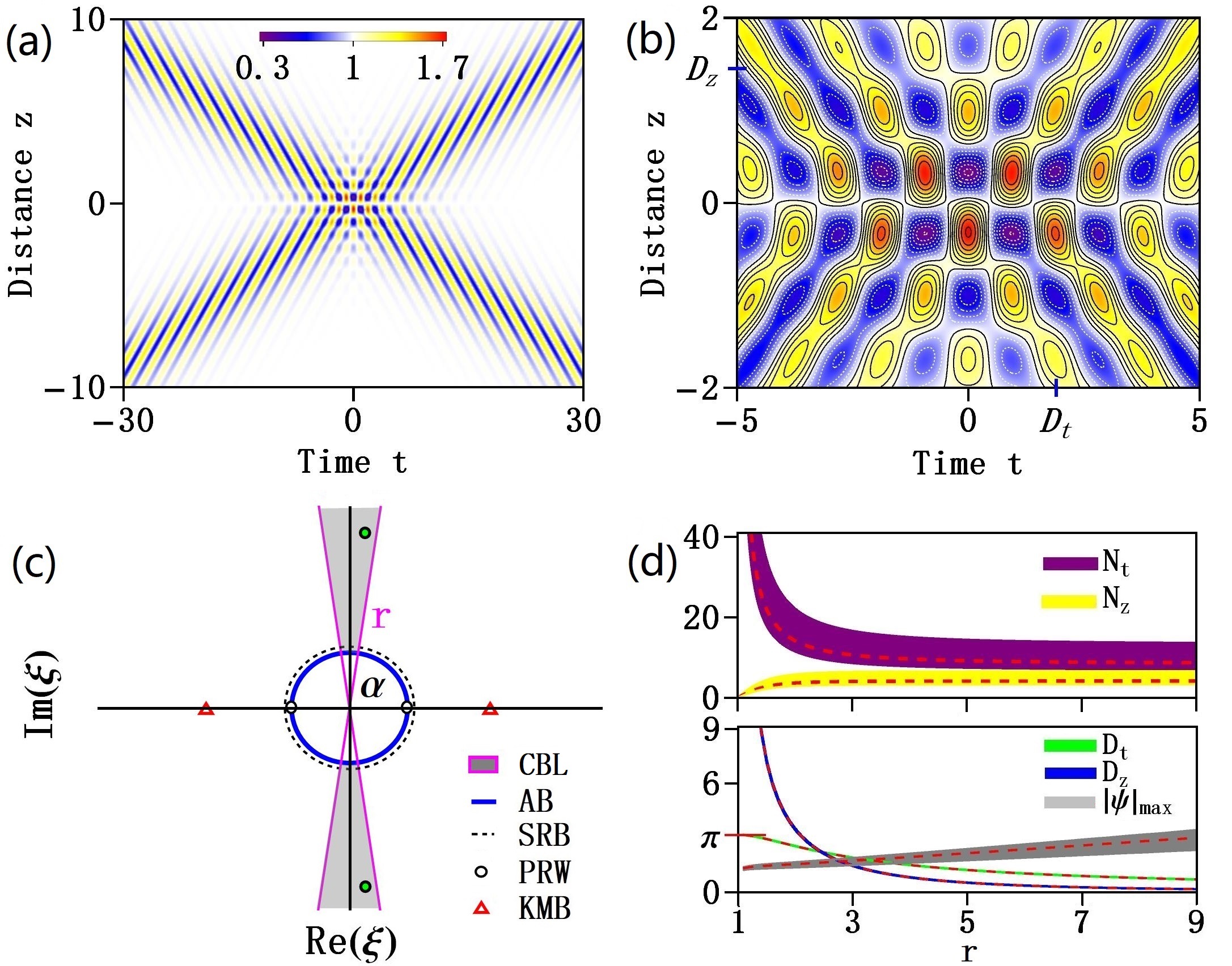}
\caption{(a) Chess-board-like interference pattern formed by the two-breather collision calculated using Eq. (\ref{eqb}) with the parameters $r=3$, $\alpha=\pi/2.16$, $\theta_1=\theta_2=\pi/2$. (b) Magnified central part of the pattern in (a).
(c) The complex plane of transformed eigenvalues $\xi=re^{\pm i\alpha}$ of two colliding breathers. Particular points or curves on this plane correspond to the following cases:  ``AB'' (Akhmediev breather, $r=1$, the unit circle), ``SRB'' (Super-regular breather, $r\rightarrow1$, $\theta_1+\theta_2=\pi$), ``PRW'' (Peregrine rogue wave, $r=1$, $\alpha=0$), and ``KMB'' (Kuznetsov-Ma breather, $r>1$, $\alpha=0$). ``CBL'' (Chess-board-like) pattern appears for all points within the triangular areas $|\alpha|\rightarrow\pi/2$ limited by the red lines outside of the blue circle $r>1$. The pair of green circles corresponds to the pattern shown in (a).
(d) Variation of $N_t$, $N_z$, interference periods ($D_t$, $D_z$), and maximum amplitudes ($|\psi|_{max}$), as $r$ changes in the range of $\alpha\in[\pi/2.2,\pi/2.1]$. The red dashed lines correspond to $\alpha=\pi/2.16$.}  \label{fig1}
\end{figure}

The solution depends on free parameters $r$, $\alpha$, and phases $\theta_j$. An example of intensity pattern produced by the above solution is shown in Fig. \ref{fig1}(a). This figure shows the collision of two breathers with chess-board-like double-periodic interference pattern in the region of collision. This pattern is qualitatively different from the soliton collision areas studied earlier. In the latter case, the interference pattern consists of parallel stripes, i.e. it is single periodic. Single periodic patterns have been also observed theoretically and experimentally in breather collision areas \cite{SR1, HB3}. In contrast, breather collisions can produce more complicated structures \cite{DJK,P1,P2}. The pattern presented in Figs. \ref{fig1}(a,b) demonstrates clearly one of these complications.
Namely, Figure \ref{fig1}(a) shows symmetric collision of two breathers propagating at finite angle to each other. For each breather, the widths in $t$ and $z$ directions are given by
\begin{equation}
\Delta t\sim1/\left|\gamma\right|,~~~\Delta z\sim1/\left|\gamma V_{gr}\right|.
\label{eqDe}
\end{equation}
The periods of breathers along $t$ and $z$ axes are
\begin{equation}\label{eqP}
D_t=2\pi/\left|\delta\right|,~~~D_z=2\pi/\left|\delta V_{ph}\right|,
\end{equation}
respectively. These values can be used to calculate the number of fringes $N_t$ and $N_z$ within the interference pattern in each direction:  $\Delta t\sim N_t D_t$, $\Delta z\sim N_z D_z$. For clear double-periodic interference pattern, both integers $N_t$ and $N_z$ should be sufficiently large. From Eqs. (\ref{eqDe}) and (\ref{eqP}), it follows that
\begin{equation}
N_t\sim\left|\tan{\alpha}\right|\left(\frac{r+1/r}{r-1/r}\right),~~N_z\sim\left|\cot{2\alpha}\right|\left(\frac{r^2-1/r^2}{r^2+1/r^2}\right).
\end{equation}
For an arbitrary $r >1$, the double-periodic interference pattern appears when $|\alpha|\rightarrow \pi/2$ (or $\left|\tan{\alpha}\right|,\left|\cot{2\alpha}\right|\rightarrow\infty$) to ensure that
$N_t$, $N_z$ are big enough.

Once $\alpha$ is fixed ($|\alpha|\rightarrow \pi/2$), the two limiting cases, (i) $r\rightarrow1$, and (ii) $r\rightarrow\infty$, can be dropped. Indeed, when $r\rightarrow1$, thus $N_t\rightarrow\infty$, $N_z\rightarrow0$, the interference pattern exhibits small amplitude oscillations only in $t$. This can be seen from Fig. \ref{fig1}(d). On the other hand, when $r\rightarrow\infty$, meaning $N_t\sim\left|\tan{\alpha}\right|$, $N_z\sim\left|\cot{2\alpha}\right|$, the interference pattern is confined to an infinitesimal region with very large amplitude. Thus, we consider interference patterns for modest values of $r$. An example of the interference pattern with $\alpha=\pi/2.16$ for $r=3$ shown in Figs. \ref{fig1}(a) and \ref{fig1}(b) exhibits double periodicity within the collision region. Its temporal and spatial periods are $D_t$, $D_z$, respectively. An interesting feature of the pattern is that there is a periodic $\pi$ phase shift at the half-periods $kD_z/2$ (where $k$ is an odd number) along the $z$ axis. This phase shift results in the chess-board-like structure.
The complex plane in Fig. \ref{fig1}(c) shows the region of existence of these patterns.
Namely, they are located in the symmetric grey-coloured triangular area on this plot. Location of eigenvalues for other types of NLSE solutions is also shown on this plot.

\begin{figure}[htb]
\centering
\includegraphics[height=32mm,width=84mm]{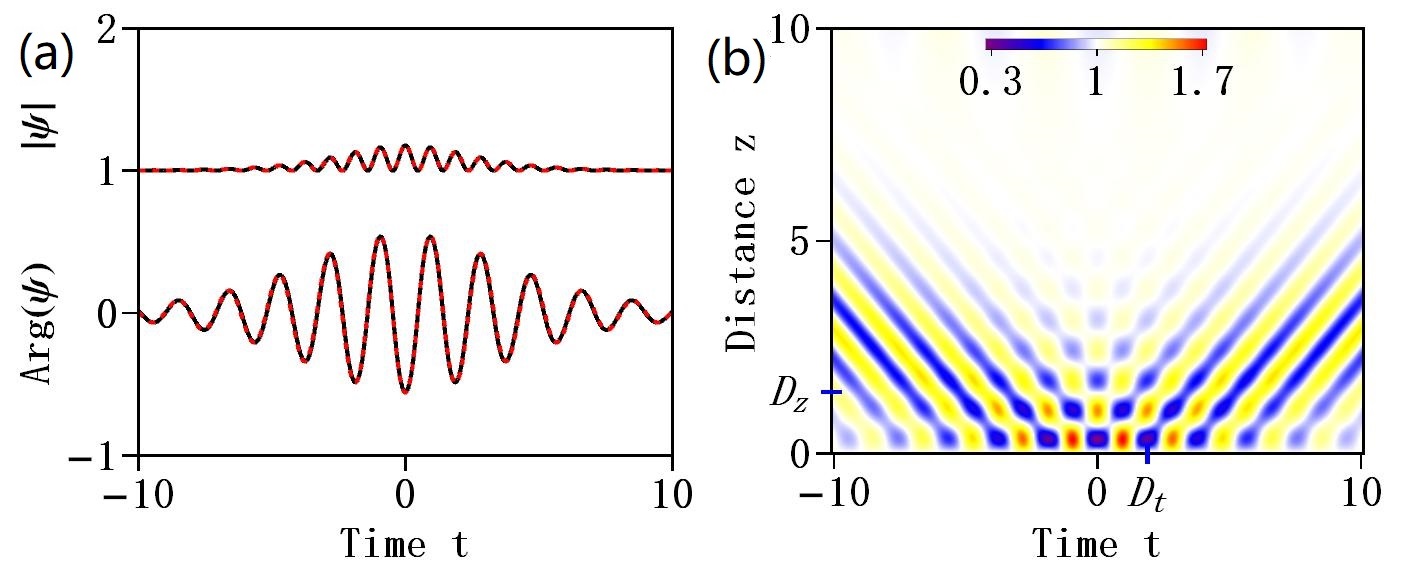}
\caption{(a) Amplitude and phase profiles of approximate initial conditions (\ref{eqi}) (red dashed line) and exact solution (\ref{eqb}) at $z=0$ (black solid line).
(b) Pattern started from the approximate condition in (a). Parameters are the same as in Fig. \ref{fig1}.
}\label{figin1}
\end{figure}

From experimental point of view and for numerical modelling, an important question is what type of initial conditions may create the chess-board-like patterns. In previous studies \cite{HB2,Extreme09,Excite09}, the breather collisions were triggered by the input field with multiple complex exponentials. Here, we extract the initial condition from the exact solution (\ref{eqb}) at $z=0$, i.e., $\psi(z=0,t)$. The phase parameters $\theta_1$, $\theta_2$ play a key role in forming the initial state. If $\theta_1+\theta_2=2k\pi$ (where $k=0,\pm1,\pm2...$), the function $H=0$. Then the initial state is purely real. If $\theta_1+\theta_2=(2k+1)\pi$, then $G=0$. Consequently, the modulation part of the initial condition is purely imaginary. General case of complex initial conditions is too complicated and does not produce anything new. Considering $|\alpha|\rightarrow\pi/2$, or $\cos^2\alpha\rightarrow0$, we analytically obtain the approximate expressions for these two types of initial conditions:
\begin{eqnarray}
\psi(0,t)_i&\approx&1-i~\rho_i~\text{sech}\gamma t\cos\bm{\delta},   \label{eqi}\\
\psi(0,t)_r&\approx&1-\rho_r~\text{sech}\gamma t\cos(\bm{\delta}+k\pi),\label{eqr1}
\end{eqnarray}
where $\rho_i=2B_1\cos\alpha/\epsilon_-$, $\rho_r=2(A_1-4\nu)\cos\alpha/(4 \cos\alpha-\epsilon_+)$, $\bm{\delta}=\delta t+\bm{\theta}$ with $\bm{\theta}=(\theta_1-\theta_2)/2$.
The approximate formulae (\ref{eqi})-(\ref{eqr1}) can be used as initial conditions for generating the interference patterns with high accuracy.
Both $\psi(0,t)_i$ and $\psi(0,t)_r$ consist of a localised function $\textrm{sech}\gamma t$ and the modulation $\cos \delta t$ responsible for the temporal period $D_t$ of interference patterns.
An example of initial conditions and the resulting pattern are shown in Fig. \ref{figin1}. There is an excellent agreement between the pattern obtained from the exact solution (\ref{eqb}) and the one simulated numerically using the initial condition (\ref{eqi}).

\begin{figure}[htb]
\centering
\includegraphics[height=32mm,width=84mm]{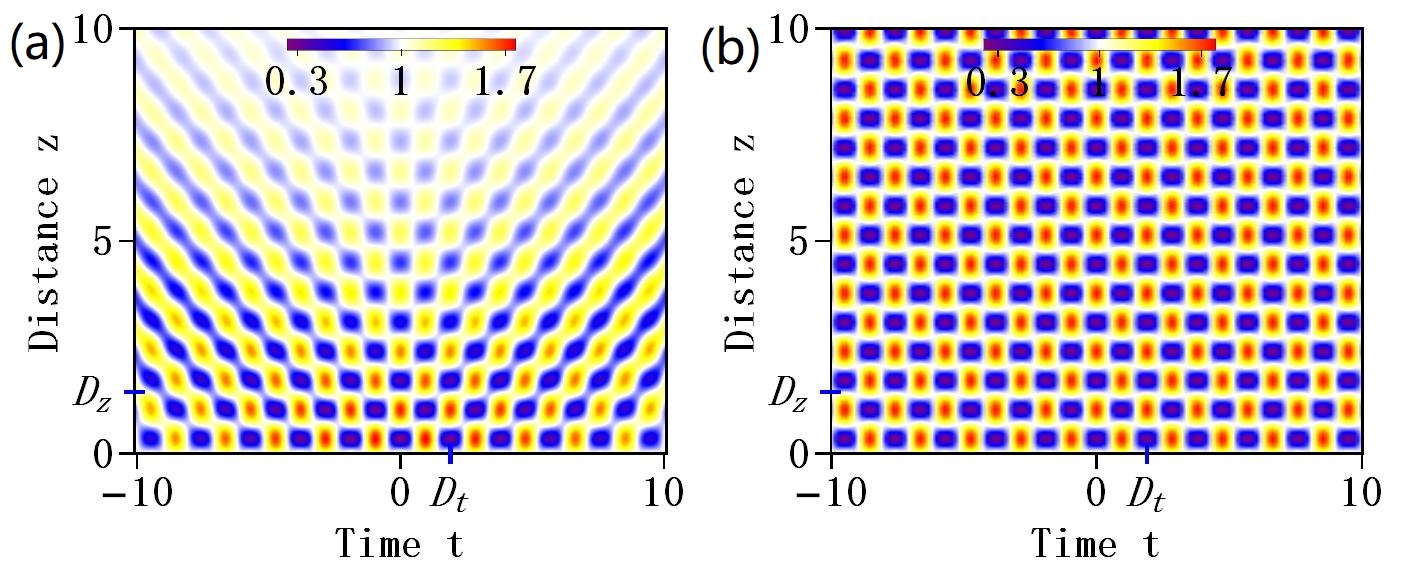}
\caption{Chess-board-like interference patterns from the initial condition $\psi(0,t)=1-i \rho_i\textrm{sech}(t/b_1)\cos{\delta t}$ with different widths (a) $b_1=3b_s$, (b) $b_1=30b_s$ where $b_s=1/|\gamma|$. Other parameters are the same as in Fig. \ref{figin1}.
}\label{figns}
\end{figure}
Clearly, the function $\psi(z,t=0)$ has the spatial period $2\pi/ |\delta V_{ph}|$. The two periods of spatio-temporal interference pattern are given by $D_t$, $D_z$.
Once $r$ and $\alpha$ are fixed, the interference patterns have these periods in $t$ and $z$ no matter what is the shape of the swell. This is because $\theta_j$ have no effect on the discrete symmetric spectra $\lambda_j$. They influence only the maximal amplitudes and location of interference maxima.
When $|\alpha|\rightarrow\pi/2$, implying that $|\sin\alpha|\rightarrow1$, the modulation frequency satisfies the condition $|\delta|=(r+1/r)|\sin\alpha|\geq2$. We can also see, from Fig. \ref{fig1}(d) that $D_t=2\pi/|\delta|\leq\pi$. This means that the chess-board-like pattern occurs in a high-frequency region out of the MI gain curve ($|\delta|<2$).

Let us take a closer look at Eqs. (\ref{eqi})-(\ref{eqr1}) by a more general case: $\psi(0,t)=1-\sigma\rho_1 \textrm{sech}(t/b_1)\cos{\delta_1t}$, where $\sigma=i$ or $1$, $\delta_1$ is the frequency, $\rho_1$ and $b_1$ are the amplitude and width, respectively.
To excite interference patterns in Fig. \ref{figin1}, $\sigma\rho_1$ can be either purely real or purely imaginary,
while the frequency must satisfy $|\delta_1|\geq2$. This frequency is outside of the MI gain curve.
For simplicity, we take $\delta_1=\delta$. The critical width of localisation which allows to excite the double-periodic pattern is $b_1=b_s=1/|\gamma|$. When $b_1\geq b_s$, numerical simulations clearly demonstrate the double periodicity. When $b_1\rightarrow\infty$, the periodic pattern occupies the whole half-plane ($z,t$). Instead, when $b_1<b_s$, the interference pattern weakens and disappears completely at small $b_1$.

According to (\ref{eqi})-(\ref{eqr1}), for any $b_1~(\geq b_s)$, purely imaginary amplitude is fixed by $\rho_1=\rho_i$; the maximal real amplitude is $\rho_1=\rho_r$. The generated interference patterns in each case have the periods $D_t$, $D_z$ defined by Eq. (\ref{eqP}).
When $b_1\rightarrow\infty$, the pattern has the same maximum amplitude $1-\rho_r$, no matter whether it is real or imaginary. In either case, the high-frequency interference patterns can be excited when $b_1\geq b_s$.  The larger $\rho_1$ is,  the bigger is the maximum amplitude of interference patterns  and the smaller is $D_z$.
Figure \ref{figns} shows an example of excitation with two different widths $b_1$. As can be seen from comparison of two panels (a) and (b), when $b_1$ increases, the size of the interference pattern gets larger while the periods $D_t$, $D_z$ remain the same. When $b_1\rightarrow\infty$, the interference pattern spreads indefinitely to occupy the whole space-time plane.

\begin{figure}[htb]
\centering
\includegraphics[height=32mm,width=84mm]{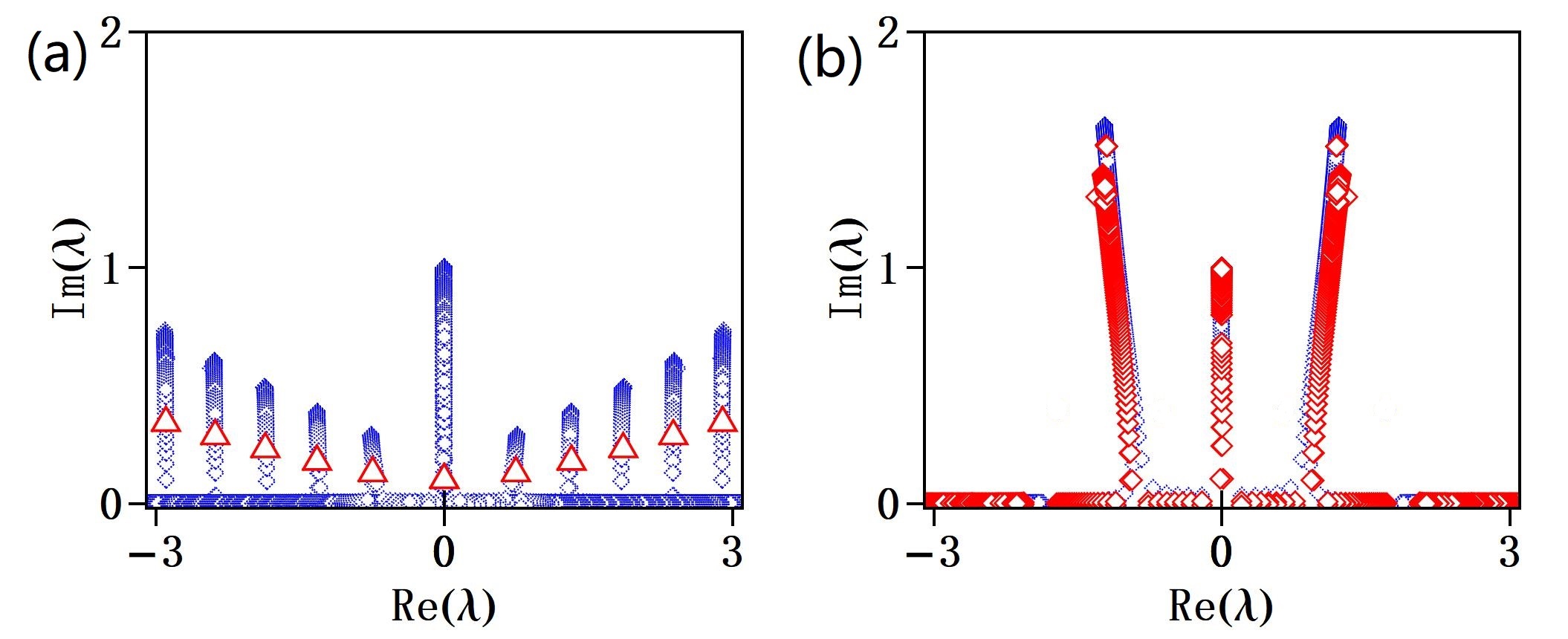}
\caption{(a) Location of discrete and continuous eigenvalues $\lambda$ on the upper complex half plane calculated for the initial condition $\psi(0,t)=1-i\rho_i\textrm{sech}(t/b_1)\cos{\delta t}$ with $b_1=b_s$ (red triangles) and $b_1\rightarrow\infty$ (blue diamonds) for $r=1,...,6$, and $\alpha=\pi/2.16$. The latter case produces six different continuous spectra. (b) Eigenvalues $\lambda$ calculated for the A-type Akhmediev solution with $\kappa=0.11$ (red) and a periodic initial condition  $\psi(0,t)=1-i 2.7\cos{3.3 t}$ (blue). The two spectra in (b) are continuous and practically overlap.
}\label{fig4-S}
\end{figure}

Initial conditions can be analysed using the eigenvalues of the NLSE theory \cite{IT3a,Yang,Randoux}. An example of such analysis is given in Fig. \ref{fig4-S}(a).
The spectrum that corresponds to the excitation of breathers is discrete. The spectrum becomes continuous in the limit of infinitely periodic solution. The latter case is similar to the double-periodic A- and B-type Akhmediev solutions of the NLSE \cite{AB1,B1,JM17}. These solutions have been studied experimentally in \cite{FPU2,FPU4}. They can be considered as a perturbation of ABs which shift solution from the heteroclinic separatrix trajectory in an infinite-dimensional  phase space to a periodic one.  These A- and B-type solutions are located on different sides of the separatrix which is the AB \cite{AB1}.
This transformation produces significant qualitative changes in the intensity pattern. Fringe pattern is transformed into the chess-board-like structure. Moreover, discrete eigenvalue spectra of breathers are transformed into continuous spectra of double-periodic solutions. In order to show this, we calculated numerically the spectrum of the A-type solution. It is shown in Fig. \ref{fig4-S}(b). The spectrum is indeed continuous with discreteness in the lower part solely due to the discreteness of numerical scheme.

The initial condition for generation of A- and B-type solutions can be expressed in terms of Jacobi elliptic functions. Their fundamental periods can be calculated exactly \cite{DP}.
Clearly, the A-type solution describes periodic structures with the transverse period located in the interval $D_t\in(0, \sqrt{2}\pi)$ whereas B-type solutions exist in the region $D_t\in(\sqrt{2}\pi, \infty)$.
The chess-board-like interference patterns can be excited when the period is located in the range $D_t\in(0, \pi)$. This means that the A-type solutions with the periods within the range $D_t\in(0, \pi)$ can describe the infinitely extended of interference pattern ($b_1\rightarrow\infty$) created by two-breather collision. Calculation of eigenvalues for the two cases confirms this. The two spectra nearly overlap as can be seen from Fig. \ref{fig4-S}(b).

More complicated patterns can be produced when using muti-swell initial conditions:
\begin{equation}
\psi(0,t)=1-\sigma\sum_{j=1}^n L_j(t-t_{j})\cos{\left(\delta_jt+\Theta_j\right)},\label{eqn}
\end{equation}
where $\sigma=i$ or $1$, $t_{j}$ is the time shift, $\delta_j$ is the frequency, $\Theta_j$ is the phase shift and $L_j(t-t_{j})$ is smooth localised functions. This can be sech-type, Gaussian,  Lorentzian or another similar function.
In what follows, we used the sech-type function, $L_j(t-t_{j})=\rho_j \textrm{sech}[(t-t_{j})/b_j]$.
When $n=1$, Eq. (\ref{eqn}) generates the two-breather interference pattern shown above. When $n\geq2$, the pattern corresponds to multiple collisions. Indeed, if the swells are well separated by the choice of $t_{j}$, Eq. (\ref{eqn}) provides a good approximation for modelling 2$n$-breather collision.

\begin{figure}[htb]
\centering
\includegraphics[height=32mm,width=84mm]{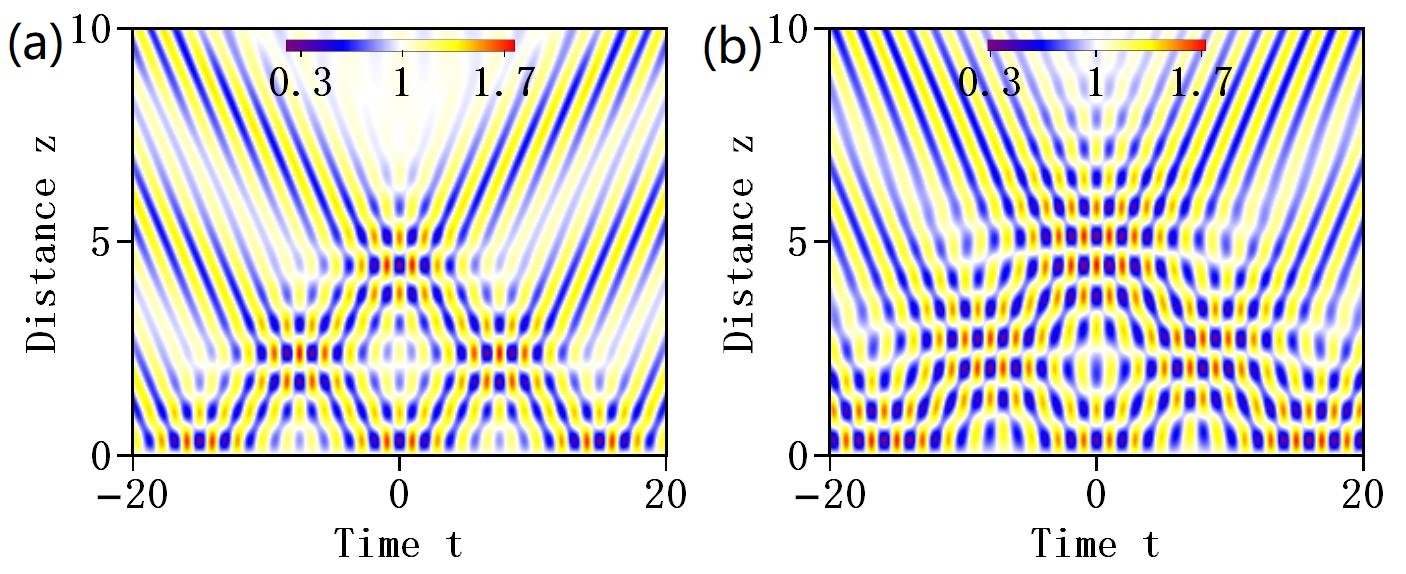}
\caption{Multiple ($n=3$) excitations of chess-board-like interference patterns started from the initial condition $\psi(0,t)=1-i\sum_{j=1}^3 \rho_i\textrm{sech}[(t-t_j)/b_j]\cos{\delta t}$ with different widths (a) $b_j=b_s$, (b) $b_j=2b_s$.
 Here $t_j=-15, 0, 15$. Other parameters are the same as in Fig. \ref{figin1}.
}\label{figin4}
\end{figure}

Figure \ref{figin4} shows two examples of excitation that started from three localised swells with the same envelope ($n=3$) with equal $b_j~(\geq b_s)$.
Each collision in Fig. \ref{figin4}(a) with $b_j=b_s$ is similar to that in Fig. \ref{figin1}(b). The total number of collisions is $n(n+1)/2$. The periods in $t$ and $z$ are well predicted by the analytical expressions for $D_t$ and $D_z$.  For larger values of $b_j$, the width of each swell becomes larger and the areas of collision also increase in size. One example with $b_j=2b_s$ is shown in Fig. \ref{figin4}(b). Overall geometry of collision is the same as in Fig. \ref{figin4}(a) but the number of high-amplitude maxima is higher.


In conclusion, we introduced new chess-board-like spatio-temporal interference patterns generated by localised periodic initial conditions applied to the NLSE. These patterns appear when the initial periodicity is outside of the MI range. We shown that they correspond to two-breather collision with specific double-periodic interference patterns. When the pattern extends to infinity, the spectrum of eigenvalues is transformed from discrete to continuous. Multiple collision patterns can be excited when the initial conditions contain several swells. Considering the universality of the NLSE, these collision-like double-periodic interference patterns can be observed in variety of physical systems ranging from optics and hydrodynamics to Bose-Einstein condensates and plasma.

This work has been done when C.L. visited Optical Sciences Group, ANU.
This work is supported by NSFC (Nos. 11705145, 11434013, and 11425522), the Major Basic Research Program of Natural Science of Shaanxi Province (No. 2017KCT-12), ARC (Discovery Projects DP140100265 and DP150102057), and the Volkswagen Stiftung. 

\end{CJK}


\begin{thebibliography}{99}
\bibitem{DP} The initial periods of A- and B-type Akhmediev solutions are
$
D_t=4\sqrt{\frac{\kappa}{2}}\textbf{K}_{m=\sqrt{\frac{1-\kappa}{2}}},~D_t=\frac{4}{\sqrt{1+\kappa}}\textbf{K}_{m=\sqrt{\frac{1-\kappa}{1+\kappa}}}$,
where $\textbf{K}_m=\int_0^{\pi/2}d\theta/\sqrt{1-m \sin^2\theta}$, and $\kappa\in(0,1)$.
The free parameter for the families of these solutions is $\kappa$ and it is assumed that the background averaged amplitude is one. From here, we find that the A-type solution describes periodic structures with the transverse period located in the interval $D_t\in(0, \sqrt{2}\pi)$ whereas B-type solutions exist in the region $D_t\in(\sqrt{2}\pi, \infty)$.
\end{thebibliography}
\end{document}